# Band Gaps and Localization in Acoustic Propagation in Water with Air-cylinders

Zhen Ye[*] and Emile Hoskinson[*,+]

*Department of Physics, National Central University, Chungli, Taiwan, ROC*

[+]Present address: Department of Physics, University of California at Berkeley, CA 94720, USA

**Multiple scattering of waves leads to many peculiar phenomena such as complete band gaps in periodic structures and wave localization in disordered media. Within a band gap excitations are evanescent; when localized they remain confined in space until dissipated. Here we report acoustic band gap and localization in a 2D system of air-cylinders in water. Exact numerical calculations reveal the unexpected result that localization is relatively independent of the precise location or organization of the scatterers. Localization occurs within a finite region of frequencies, coincident with the complete band gap predicted by a conventional band structure calculation for a periodic lattice of scatterers. Inside the gap or localization regime, a previously uninvestigated stable collective behavior of the cylinders appears.**

Multiple scattering of waves is responsible for a wide range of fascinating phenomena[1,2]. This includes twinkling light in the evening sky, modulation of ambient sound at the ocean surfaces[3], and acoustic scintillation of turbulence[4]. On smaller scales, phenomena such as random lasers[5] and electron transport in impure solids[6] are also explained by multiple scattering. Under appropriate conditions, multiple scattering leads to photonic or acoustic band gaps in classical wave propagation in periodic structures[7,8], and to the unusual phenomenon of wave localization, a concept introduced by Anderson[9] to explain the conductor-insulator transition induced by disorders in electronic systems.

In periodic structures, the wave dispersion bands are well understood in the realm of Bloch's theorem and have been observed for optical systems[7] and acoustic systems[8]. In random scattering media, however, a peculiar phenomenon of wave localization may occur. In disordered solids, electron localization is common[10]. Random underwater topography can give rise to water wave localization[11]. Localization effects have also been reported for microwaves[12,13], for light[14], and for acoustic waves[15,16,17,18] in disordered media. Here we report acoustic localization in the strongly scattering 2D random arrays of air-cylinders in water, with comparison to the complete acoustic band gap of the corresponding regular arrays. We show that acoustic localization can be achieved in a range of frequencies, coincident with the complete band gap. Within the band gap or localized region, the system exhibits an interesting coherent behavior. A new method is proposed to investigate the phase transition between localized and extended states. The approach enables to isolate the localization effect from that due to residual absorption.

Multiple scattering of waves is established by an infinite recursive pattern of rescattering between scatterers. In terms of wave field $u$, The energy flow in the system is calculated from $\mathbf{J} \sim Re[u(\sqrt{-1})\nabla u]$. Writing $u = |u|e^{i\theta}$, the flow becomes $\mathbf{J} \sim |u|^2 \nabla\theta$, a Meissner equation. Obviously, the energy will be localized when phase $\theta$ is a constant and $|u| \neq 0$. This simple perception will guide our following discussion.

The system is constructed by placing $N$ uniform air-cylinders of radius $a$ in water perpendicular to the $x-y$ plane to form either a square lattice with lattice constant $d$ or a random array. Experimentally, the air-cylinders can be gas enclosures shelled by a thin insignificant elastic layer, like the Albunex® bubbles used in echocardiography[19]. A unit acoustic pulsating line source transmitting monochromatic waves is placed at the center of the array. Due to the large contrast in acoustic impedance between air and water, the air-cylinders are strong acoustic scatterers. The scattered wave from each cylinder is a response to the total wave incident upon it. The total incident wave is composed of the direct contribution from the source and the multiply scattered waves from each of the other cylinders. The response function of a single cylinder is readily computed by a modal-series expansion in cylindrical coordinates, in the form of the partial wave solution, by invoking the usual boundary conditions that the pressure and the radial velocity be continuous across the cylinder surface. Unlike the 3D bubble case, absorption caused by thermal exchange and viscosity is unimportant[20] and is not included in the model. Strong single cylinder scattering occurs for $ka$ ranging from roughly $0.0005$ to $0.5$ with a resonant peak at $ka \approx 0.006$. Here $k$ is the wavenumber and $a$ the cylinder radius. In this range of $ka$, to which our attention is restricted, the radial pulsating vibration of the cylinders provides the main contribution to the scattering. Multiple scattering in such a system can be calculated *exactly*, using a self-consistent method[21] and a matrix inversion scheme[18].

Each cylinder acts as an effective secondary pulsating line source. The scattered wave from the $i$-th cylinder can be written as $A_i H_0^{(1)}(k|\mathbf{r}-\mathbf{r}_i|)$, where $H_0^{(1)}$ is the zero-th order Hankel function of the first kind and $\mathbf{r}_i$ denotes the position of the cylinder. The complex coefficients $A_i$ represent the effective source strength, and are computed incorporating *all orders of* multiple scattering. The total wave at any space point is the sum of the direct wave from the transmitting source and the scattered wave from all the cylinders.

We express $A_i$ as $|A_i|\exp(j\theta_i)$, with $j=\sqrt{-1}$ and $i=1,2,...,N$. The modulus $|A_i|$ represents the strength and $\theta_i$ the oscillation phase of the $i$-th scatterer relative to the actual source. A unit vector $\mathbf{u}_i$, hereafter termed the phase vector, is associated with each phase $\theta_i$, and these vectors are represented on a phase diagram in the $x-y$ plane.

The starting point of each phase vector is positioned at the center of its corresponding cylinder and oriented at an angle with respect to the positive $x$-axis equal to the phase: $\mathbf{u}_i = \cos\theta_i \mathbf{e}_x + \sin\theta_i \mathbf{e}_y$. Numerical experiments are carried out to study the behavior of the phases of the cylinders and the spatial distribution of the acoustic energy. The fraction of area occupied by the cylinders per unit area $\beta$ is set to $10^{-3}$, and the total number of scatterers $N$ is at least 200. The lattice parameter $d$, or the average distance between neighboring cylinders in the random case, is calculated as $d \sim (\pi/\beta)^{1/2} a$.

Figure 1 presents the far field transmitted intensity as a fraction of what it would in the absence of scatterers. It is plotted as a function of frequency in terms of the dimensionless parameter $ka$ for an ordered square lattice, and for a random array of the scatterers. Inside a finite frequency range, transmission is significantly inhibited for both cases; this inhibition indicates that waves cannot escape, rather they are trapped in the vicinity of the source. Interestingly, the presence and qualitative features of the localization regime are not changed between the ordered and random configurations. The main effect of the added disorder is to smooth out the upper edge of the localized regime. We have observed that increasing or decreasing $\beta$, which decreases or increases the nearest-neighbor separation, causes this edge to move upwards or downwards respectively. It is likely then that the smoothing effect observed in the random configuration is due to the introduction of a distribution in nearest neighbor separations.

We can compare the transmission inhibition behavior, in the ordered case, with a conventional band structure calculation. The band structure is shown in Figure 2. We observe the presence of a complete acoustic band gap between $ka = 0.007$ and $0.03$, which matches with the transmission inhibition regime indicated in Figure 1. In this region, the system has no propagation modes in any direction. It is also observed that if $\beta$ is reduced, the gap narrows.

We further explore how the system behaves for frequencies below, within and above the gap and localization regimes, in connection to the phase vectors. Figures 3 and 4 show respectively the resulting phase vectors and the energy distribution for three frequencies for both regular and random arrays.

For frequencies below the gap or localization regime, there is no obvious ordering for the phase vectors $\mathbf{u}_i$, nor for the energy distribution. The case of $ka = 0.0048$ is representative; the phase vectors point to various directions, and no energy localization appears. In the ordered case, this behavior is due to boundary effects. In effect, as the wave is not localized, it can travel through the array and is reflected by the cylinders near the asymmetric border; all cylinders experience the effect through strong multiple scattering. In the random arrays, not only does the boundary play a role, but the random placement of cylinders also contributes to the random phase and energy distribution.

As the frequency increases into the localized or gap regime, an ordering among the phase vectors indeed becomes evident. This is shown for $ka = 0.01$ for both random and ordered arrays. Amazingly, all cylinders oscillate completely in phase, but exactly out of phase with the transmitting source. Such collective behavior allows for efficient cancellation of incoming waves. Both the regular and random cases clearly show that energy is trapped or localized near the source, decaying exponentially with distance from it. The localization length, defined as the inverse of the decay rate in the random case, increases steadily as $ka$ increases through the localized regime, causing the transmission in Figure 1 to increase towards the upper regime edge. At $ka = 0.01$, the localization length is about $30a$. The exponential energy decay occurs for frequencies within the complete band gap of the corresponding ordered case. The wave localization is independent of the outer boundary and always appears for sufficiently large $\beta$ and $N$. Such localization and phase behavior is related to the particular scattering properties of the air-cylinders. In the context of field theory[22], the global collective phenomenon implies a symmetry breaking and appearance of a certain kind of Goldstone mode.

When the frequency is increased beyond the upper gap or localized regime edge, the in-phase ordering disappears. Meanwhile, the wave becomes extended again. This is illustrated by the case of $ka = 0.035$. Although the cylinders do not oscillate in phase, there is a periodic pattern in the phase vectors for the regular case: the cylinders at equal distance from the source have the same phase vectors. This is explained by weak multiple scattering. When multiple scattering is weak, the cylinders are mainly driven by the direct incident wave from the source; therefore the phase $\theta_i$ of the $i$-th cylinder approximates $k|\mathbf{r}_i|$ plus a constant phase from the response function of the cylinder. The pattern reflects the periodic arrangement of the cylinders, and the energy distribution has a matching periodicity.

We emphasize that the localization features in the random array case are not due to the finite size or boundaries of the scatterer arrays. When the sample size is increased while the cylinder concentration is kept constant, inside the localization regime the exponential energy decay continues at an unchanged rate, while the wave remains non-localized for frequencies outside the localization range. Moreover, the cylinder scattering strength is actually stronger at $ka = 0.005$ than it is at $ka = 0.01$. If the behavior of the system were continuous from $ka = 0.01$ to $0.005$, according to theories we would expect the stronger scattering strength at $ka = 0.005$ to result in a stronger localization. This does not occur. In effect, the phase diagrams in Figure 4 clearly indicate that the states at $ka = 0.005$, $ka = 0.01$ and $ka = 0.035$ belong to different phase states of the system.

We have demonstrated a new localization phase transition in acoustic propagation in the 2D random distribution of uniform air cylinders and their connection to the complete band gaps in the corresponding lattice arrangements, and we suggest that such a transition is related to the collective behavior of the scatterers. These results are special, in that they apply to the particular case of isotropic resonant scatterers, and also surprising, in that they are insensitive to details of the spacing of the scatterers. Although these properties may not hold in general, the fact that they do for resonant air-cylinders makes these scatterers ideal for theoretical and experimental localization studies.

Acknowledgment. Haoran Hsu is thanked for discussion and programming assistance. The work received support from NCU and the National Science Council.

**Correspondence and request for materials should be addressed to Z.Y. (e-mail: zhen@joule.phy.ncu.edu.tw)**

Figure 1. Acoustic bands for a square lattice of air-cylinders in water. The frequency is expressed in terms of non-dimensional $ka$. The complete band gap lies between the two horizontal dotted lines.

Figure 2. Acoustic transmission vs. frequency ($ka$) for three sample sizes. The results are normalized such that the transmission equals unity when no scatterers are present.

Figure 3: Left column: Phase diagrams for the two-dimensional phase vectors $\mathbf{u}_i$ lying on the $x-y$ plane. Right column: Spatial distribution of acoustic energy (arbitrary scale). Here the $x$ and $y$ axes give the cylindrical positions in units of the cylinder radius $a$. The $z$-axis of the energy distribution plots gives the energy density in log scale relative to what it would be for an unscattered source in empty space. Plotting the relative energy density in this way removes uninteresting geometric spreading.

Figure 4: The phase diagram and energy distribution for the case of a random array of air-cylinders in water. The legends are the same as in Fig.3.

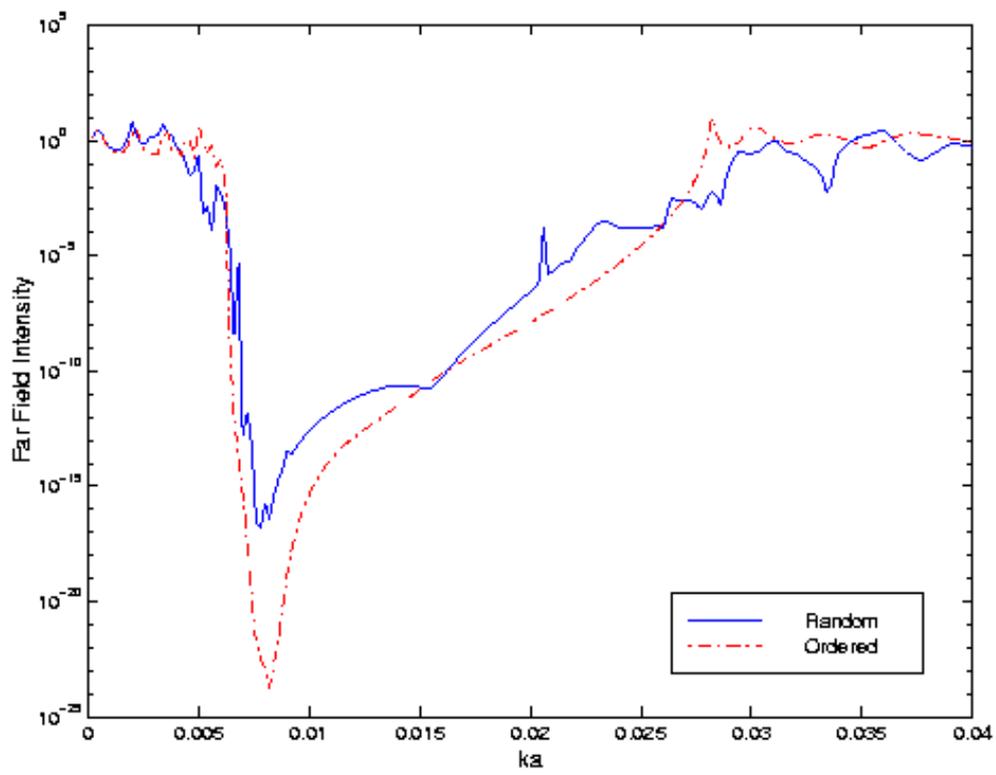

Figure 1

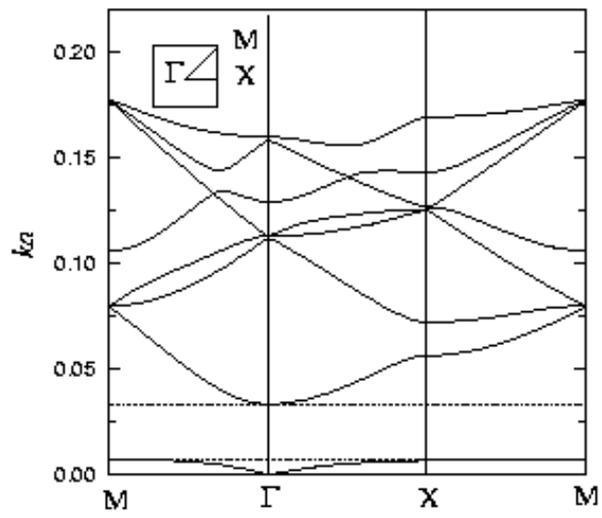

Figure 2

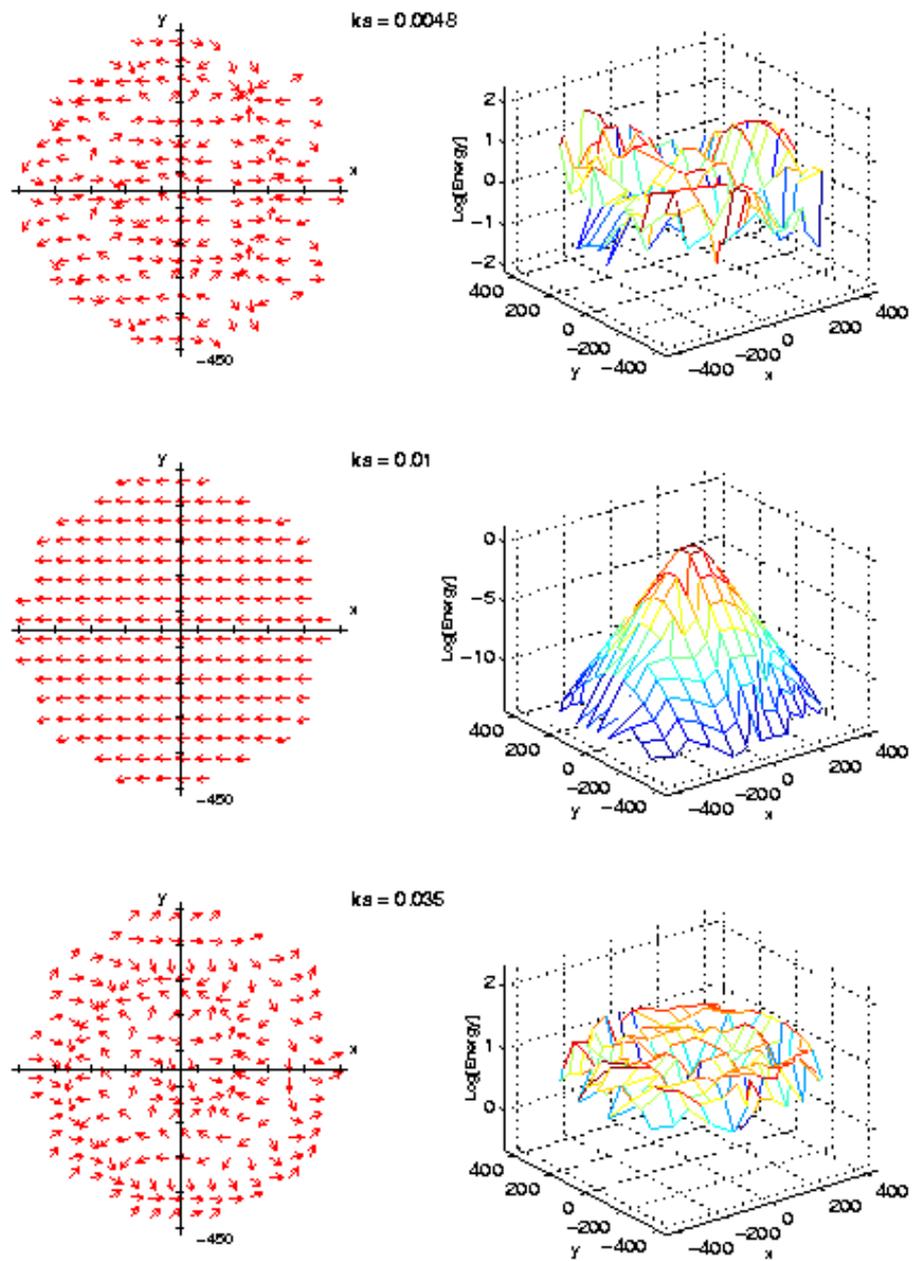

Figure 3

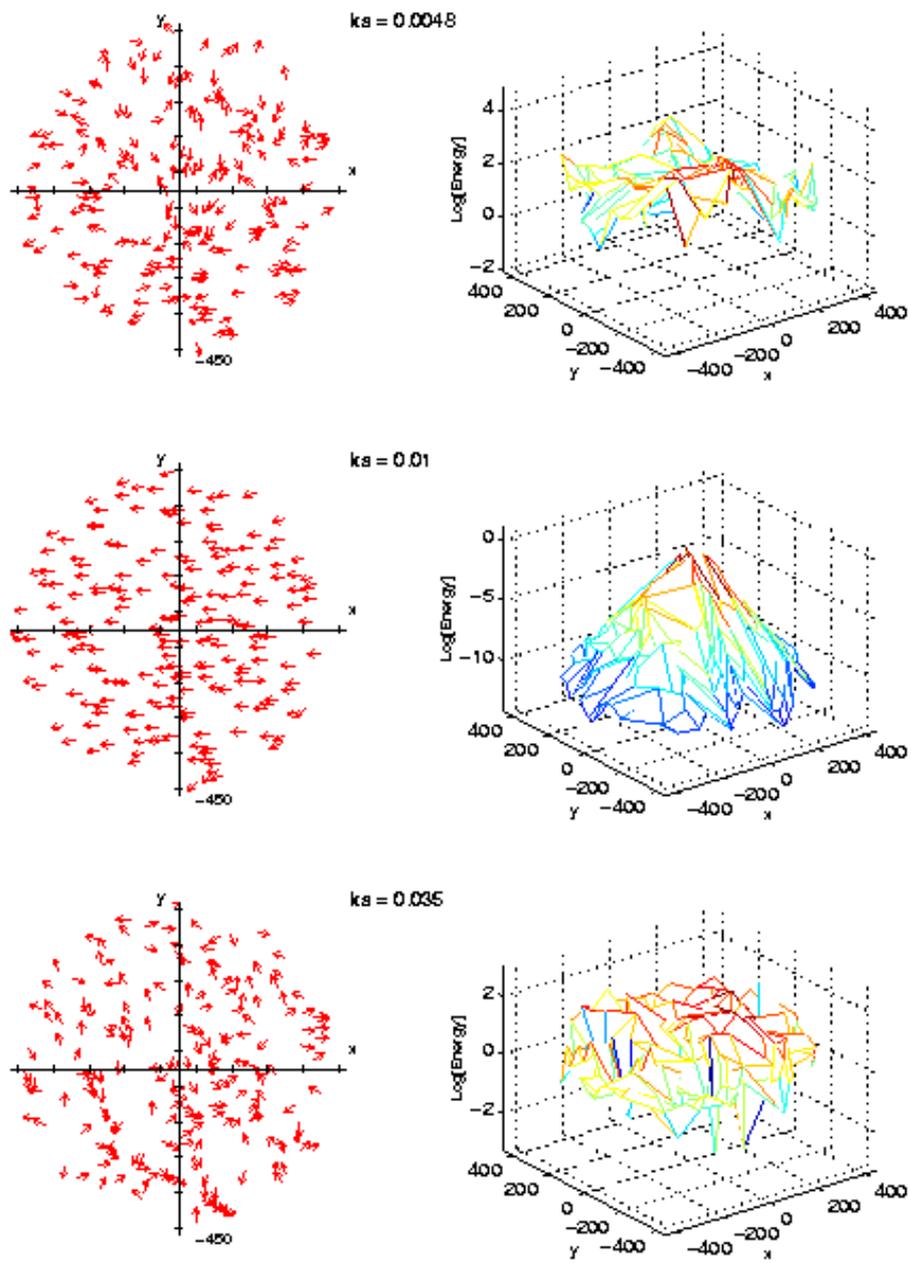

Figure 4